\documentstyle[preprint,revtex]{aps}
\begin{document}
\thispagestyle{empty}
\font\fortssbx=cmssbx10 scaled \magstep2
\hbox{ %$\vcenter{\special{insert $disk1:[pheno.tex.inputs]uwlogo.imp}}$
%\hskip.2in $\vcenter{
\fortssbx University of Wisconsin - Madison} %$ }
\vspace{.3in}
\hfill\vbox{\hbox{\bf MAD/PH/693}
            \hbox{\bf FERMILAB-PUB-92/61-T}
            \hbox{February 1992}}\par
\vspace{.2in}
\begin{title}
Test of the Dimopoulos-Hall-Raby Ansatz\\ for Fermion Mass Matrices
\end{title}
\author{V.~Barger,$^*$  M.~S.~Berger,$^*$ T.~Han$^{**}$ and
M.~Zralek$^\dagger$}
\begin{instit}
$^*$Physics Department, University of Wisconsin, Madison, WI 53706, USA\\
$^{**}$Fermi National Accelerator Laboratory, P.O.~Box 500, Batavia,
IL 60510, USA\\
$^\dagger$Physics Department, University of Silesia, Katowice, Poland
\end{instit}
\begin{abstract}
\nonum\section{abstract}
By evolution of fermion mass matrices of the Fritzsch and
the Georgi-Jarlskog forms from the supersymmetric grand unified
scale, DHR obtained predictions for the quark masses and
mixings. Using Monte Carlo methods we test these predictions
against the latest determinations of the mixings,  the CP-violating
parameter $\epsilon_K^{}$ and the $B_d^0$-$\bar B_d^0$ mixing parameter $r_d$.
The acceptable solutions closely specify the quark masses and mixings,
but lie at the edges of allowed regions at 90\% confidence level.
\end{abstract}

\newpage
One of the outstanding problems in particle physics is that of explaining the
fermion masses and mixings. In the Standard Model (SM) the 6~quark masses,
3~charged lepton masses, the 3~quark mixings and the CP-violating phase of the
Cabibbo-Kobayashi-Maskawa (CKM)
matrix are introduced as phenomenological parameters.
Over the years various models have been proposed to reduce the
number of these free parameters~\cite{one}, of which the best known is the
Fritzsch model~\cite{fritz}. Recently Dimopoulos, Hall and Raby
(DHR) have proposed an ansatz for fermion mass matrices~\cite{DHR} in the
framework of minimal
supersymmetric (SUSY) Grand Unified Theories (GUT). The DHR approach is
based on the observation that some discrete symmetries present at the grand
unification scale are broken in the low-energy theory. Thus some elements of
the fermion mass matrices that vanish at the GUT scale are non-zero at the
electroweak scale, and their low-energy values are calculable from the
renormalization group equations. The fermion masses and mixings at the
electroweak scale can thereby be expressed in terms of a smaller number of
input parameters at the GUT scale. DHR work in the massless neutrino limit and
relate the 13 SM parameters and a SUSY parameter $\tan\beta$ (discussed below)
to 8~input parameters, leading to
6~predictions that include an allowed range of 147--187~GeV for the top
quark mass ($m_t$). In comparison the Fritzsch approach gives
$77\le m_t\le 96$~GeV~\cite{one}, which is nearly excluded
in the SM by the CDF experiment~\cite{cdf} at a 90\% confidence level (C.L.).

The DHR quark mass matrices at the scale $m_t$ are
\begin{equation}
{\cal M}_u = \left( \begin{array}{c@{\quad}c@{\quad}c}
0 & C & 0 \\ C & \delta_u & B \\ 0 & B & A
\end{array} \right) {v\sin\beta\over\sqrt2} \qquad
{\cal M}_d = \left( \begin{array}{c@{\quad}c@{\quad}c}
0 & Fe^{i\phi} & 0 \\ Fe^{-i\phi} & E & \delta_d \\ 0 & 0 & D
\end{array} \right) {v\cos\beta\over\sqrt2} \;,
\end{equation}
where all the parameters are real, $\tan\beta=v_2/v_1$ in terms of the Higgs
doublet vacuum expectation values, and $v=246$~GeV. The charged lepton mass
matrix ${\cal M}_e$ is obtained from the above form of ${\cal M}_d$ by the
substitutions $\phi=0,\ \delta_d=0,\ E \rightarrow -3E',\ D\to D',\ F\to F'$.
At the SUSY-GUT scale, the parameters $\delta_u$ and $\delta_d$ vanish and
$D=D',\ E=E',\ F=F'$,
so the input mass matrix ${\cal M}_u$ is of the Fritzsch form~\cite{fritz}
and ${\cal M}_d$ and ${\cal M}_e$
are of the Georgi-Jarlskog form~\cite{G-J}, giving the GUT
scale mass relations $m_b=m_\tau,\ m_s\simeq m_\mu/3,\ m_d\simeq 3m_e$ between
quarks and leptons. The mass ratio prediction
\begin{equation}
(m_d/m_s)(1-m_d/m_s)^{-2} = 9(m_e/m_\mu)(1-m_e/m_\mu)^{-2}
\label{mass ratio}
\end{equation}
holds at all scales.

The Wolfenstein parameterization~\cite{wolf} of the
CKM matrix determined from the unitary matrices that diagonalize the DHR mass
matrices  can be expressed in terms of four
angles ($\theta_i$) and a complex phase ($\phi$) as follows
\begin{mathletters}
\begin{eqnarray}
&\lambda = (s_1^2+s_2^2+2s_1s_2\cos \phi)^{1\over 2}
= |V_{cd}| = |V_{us}| \;,\label{lambda}\\
&\lambda ^2A  = s_3-s_4  = |V_{cb}| \;, \\
&\lambda\sqrt{\rho ^2+\eta ^2}  = s_2
= |V_{ub}/V_{cb}|=\sqrt{m_u/m_c} \:, \\
&\eta  = s_1s_2\sin \phi/\lambda ^2 \;,
\end{eqnarray}
\end{mathletters}
with $s_i=\sin\theta_i, \ c_i=\cos\theta_i \  (i=1,2,3,4)$,
where $\theta_2$, $\theta_3$ are the angles that diagonalize the
matrix ${\cal M}_u$, and $\theta_1$, $\theta_4$ are those for
${\cal M}_d$~\cite{DHR}; only three of these angles are independent.
These mixing angles are related to the quark masses and other parameters by
\begin{equation}
s_1\simeq\sqrt{m_d/m_s}\,, \quad s_2\simeq\sqrt{m_u/m_c}\,,
\quad s_3\simeq|B/A|\,, \quad s_4 \simeq s_3-|V_{cb}| \,.\label{s_i}
\end{equation}

The evolution based on the SUSY renormalization group equations
(RGE) from the GUT scale to the
appropriate fermion mass scales, taking all SUSY particles and the
second Higgs doublet degenerate at the
scale of $m_t$ \cite{DHR}, gives the following relations,
\begin{mathletters}
\begin{eqnarray}
&\displaystyle
m_t = {m_bm_c\over m_\tau|V_{cb}|^2 } \, {x\over\eta_b\eta_c\eta^{1/2}} \;,
\qquad\quad m_s-m_d = \case{1}/{3} m_\mu\eta_s\eta^{1/2}/x \;, \label{s-d}\\
%\label{m_t}
&\displaystyle
\sin\beta = {m_t\over\pi v}\sqrt{3I\over2\eta}
\left[ 1-y^{12} \right]^{-1/2}\,, \qquad\quad
s_3 = {|V_{cb}|m_b\over\eta^{1/2}\eta_b m_\tau}x \;, \label{s_3}
%\label{sinbeta}
\end{eqnarray}
\end{mathletters}
where
\begin{mathletters}
\begin{eqnarray}
% \eta &=& \prod _i\left ({{\alpha_G^{}}\over {\alpha _i}}\right )^{{c_i}
%\over {b_i}}\;, \\
% I &=& \int_{m_t}^{M_G^{}}\eta (\mu)d\ln \mu \;, \\
& x = (\alpha_G^{}/\alpha_1)^{1/6} (\alpha_G^{}/\alpha_2)^{3/2} \;, \qquad\quad
 y = x(m_b/m_\tau) \eta^{-1/2} \eta_b^{-1} \;, & \\
 & \displaystyle \eta(\mu) = \prod_{i=1,2,3}(\alpha_G^{}/\alpha_i)^{c_i/b_i}
\;,
\qquad\quad
 I(\mu) = \int\limits_\mu^{M_G^{}}\eta(\mu') d\ln\mu' \;.
\end{eqnarray}
\end{mathletters}
The RGE parameters $b_i\,,\,c_i$ are given in Ref.~\cite{DHR}. In these
equations
the couplings $\alpha_1$ and $\alpha_2$ are evaluated at the scale $m_t$.
The mass parameters are defined as $m_q(\mu=m_q)$ for quarks heavier
than 1~GeV, and the lighter quark masses  $m_s,\,m_d,\,m_u$
are calculated at the scale $\mu = 1$~GeV.

Starting from  the well-determined values~\cite{amaldi},
$\alpha_1(M_Z)=0.016887,\; \alpha_2(M_Z)=0.03322$,
and evolving at one-loop level to their intersection determines the GUT
scale $M_G^{}=1.1\times 10^{16}$ GeV and the GUT coupling constant
$\alpha_G^{}=1/25.4$.
Evolving backwards, the strong coupling constant
$\alpha_s(M_Z) = 0.106$ is obtained,
consistent with the LEP result $\alpha_s=0.118\pm0.008$~\cite{carter}.
Also the values $\alpha_1(m_t) = 0.017$ and $\alpha_2(m_t) = 0.033$
are determined,  as well as the factors $\eta(m_t) = 9.7$ and $I(m_t) = 110$.
We have used a top-quark threshold of 180~GeV in the
RGE, consistent with our output determination.
In evolution below the electroweak scale we include 3-loop QCD
and 1-loop QED effects in the running masses to obtain the evolution factors
$\eta_b=1.44,\ \eta_c=1.80$ and $\eta_s=1.95 $ where
$\eta_q=m_q(m_q)/m_q(m_t)$ for $q=b,c$ and
$\eta_s=\left[m_s({\rm 1\,GeV})/m_{\mu}({\rm 1\,GeV})\right]
\Big/ \left[m_s(m_t) /m_{\mu}(m_t)\right]$.
Quark and lepton thresholds were handled by demanding
that the couplings and running masses be continuous. The number of active
flavors in the $\beta$-functions and in the anomalous dimensions was changed as
 each successive fermion was integrated out of the theory.

Following DHR, we take the following 8 relatively better-known parameters as
inputs:
$m_e,\ m_\mu,\ m_\tau,\ m_c,\ m_b,\ m_u/m_d,\ |V_{cb}|$ and $|V_{cd}|$.
We generate random values for all inputs within 90\%~C.L.
($1.64\sigma$) ranges. The input mass values~\cite{data,mass} are
$m_\tau = 1784.1 \raisebox{-1ex}{$\stackrel{\displaystyle{}+2.7}{{}-3.6}$}\,
{\rm MeV},\ m_c(m_c) = 1.27\pm0.05 \,{\rm GeV},\ m_b(m_b) = 4.25\pm0.1\,{\rm
GeV}$, where $1\sigma$ errors are quoted.
We also impose the theoretical constraint
$0.2 \le m_u/m_d \le 0.7$~\cite{ratio}.
We next calculate $m_d,\ m_s$ from Eqs.~(\ref{mass ratio}) and (\ref{s-d})
(obtaining $m_d=5.93$~MeV, $m_s=146.5$~MeV),
$s_1$ and $s_2$ from Eq.~(\ref{s_i}), $s_3$ from Eq.~(\ref{s_3}) for the
input of $|V_{cb}|$, $s_4$ from Eq.~(\ref{s_i}),
and $\phi$ from $|V_{cd}|$ of Eq.~(\ref{lambda}).
Using these values we evaluate the magnitudes of all elements
of the CKM matrix. We retain only those Monte Carlo events that satisfy the
following ranges from the 1992 Review of Particle
Properties~\cite{data},
\begin{equation}
|V_{\rm CKM}| = \left(\begin{array}{c@{\quad}c@{\quad}c}
0.9747\mbox{--}0.9759 & 0.218\mbox{--}0.224 & 0.002\mbox{--}0.007\\
0.218\mbox{--}0.224 & 0.9735\mbox{--}0.9751 & 0.032\mbox{--}0.054\\
0.003\mbox{--}0.018 & 0.030\mbox{--}0.054 & 0.9985\mbox{--}0.9995
\end{array}\right) \;,
\end{equation}
as well as the ratio $0.051 \le |V_{ub}/V_{cb}| \le 0.149 $~\cite{data}.
Figure~\ref{scatter} shows a scatter plot of $\sin\beta$ versus $m_t$
obtained from our Monte Carlo analysis; one sees that only a narrow wedge of
the space is permissible.

{}From $\sin\beta\le 1$ in Eq.~(5), $|V_{cb}|$ must satisfy the inequality
\begin{equation}
|V_{cb}| \agt \left[ {x\over\pi v} \sqrt{ 3I\over  2(1-y^{12}) } \,
{m_bm_c\over m_\tau \eta \eta_b\eta_c} \right]^{1/2} \agt 0.053 \;,
\end{equation}
which is just at the edge of the 90\% C.L. allowed range.
Calculating $m_t$ from Eq.~(\ref{s-d}) and $\sin\beta$ from
Eq.~(\ref{s_3}) and requiring that $|V_{cb}|$ be
within its allowed range, we find
%
%\begin{mathletters}
\begin{eqnarray}
& 174 < m_t < 183\,\rm GeV \;, \qquad
 \sin\beta > 0.954 \ (\tan\beta>3.2) \;. &
\end{eqnarray}
%\end{mathletters}
%
This top quark mass determination is consistent with estimates
from the electroweak radiative corrections~\cite{carter}
but is much more restrictive. The predicted value of $\tan\beta$
is large, which may have significant phenomenological
implications for Higgs boson searches at colliders~\cite{vb}.

Next we include the constraints from the measured values
\begin{equation}
|\epsilon_K^{}| = (2.259\pm 0.018)\times 10^{-3}  \ \cite{data} \;, \qquad
r_d = 0.181\pm 0.043\ \cite{bbbar} \;,
\end{equation}
of the CP-violating parameter $\epsilon_K^{}$ and the $B_d^0$-$\bar B_d^0$
mixing
parameter $r_d$,
The theoretical formulas, including QCD corrections, can be found in Eqs.~(2.1)
and (2.10) of Ref.~\cite{joanne}.
In our Monte Carlo analysis we allow variations of
the bag-factors and $B$-decay constant over the following ranges~\cite{joanne}:
\begin{equation}
0.33 \le B_K \le 1.5 \;, \qquad
 0.1{\rm\,GeV}\le\sqrt{B_B\,} f_B < 0.2\,\rm GeV \;,
\end{equation}
taking $f_K=160$~MeV and $\Delta M_K^{}=3.521\times 10^{-15}$~GeV.
The solutions so obtained closely
specify the CKM matrix to be
\begin{equation}
|V_{\rm CKM}| = \left(\begin{array}{c@{\quad}c@{\quad}c}
0.9749\mbox{--}0.9759 & 0.2185\mbox{--}0.2230 & 0.0027\mbox{--}0.0032\\
0.2185\mbox{--}0.2230 & 0.9735\mbox{--}0.9745 & 0.0530\mbox{--}0.0540\\
0.0106\mbox{--}0.0109 & 0.0518\mbox{--}0.0529 & 0.9985\mbox{--}0.9986
\end{array}\right) \;,
\end{equation}
and predict the CP-violating phase to be in the range
\begin{equation}
70^{\circ} < \phi < 80^{\circ}.
\end{equation}
The inclusion of $\epsilon_K^{}$  and $r_d$ almost uniquely determines the
values of $|V_{td}|$ and $|V_{ts}|$.
Since $|V_{cb}|$ is near its allowed upper limit, $|V_{ub}|$ is pushed to its
lower end by the unitarity condition. The output value of the ratio
\begin{equation}
0.051 < |V_{ub}/V_{cb}| < 0.059
\end{equation}
is at the low end of the allowed range. Improved experimental
determinations of $|V_{cb}|$ and $|V_{ub}/V_{cb}|$ will
test the DHR ansatz.
In terms of the Wolfenstein parameters~\cite{wolf}, we find
%
%\begin{mathletters}
\begin{equation}
\begin{array}{rcl}
&0.2185 < \lambda < 0.2230 \;, \qquad 1.07 < A < 1.13 \;, &\\
&0.195< \eta < 0.243 \;, \quad  0.105< \rho < 0.129 \;,
\quad  0.222< \sqrt {\rho^2 + \eta^2} < 0.275 \;.&
\end{array}
\end{equation}
%\end{mathletters}
%

The output values of the mass ratios of the light quarks are
\begin{equation}
0.52 < m_u/m_d < 0.70 \;, \quad m_d/m_s = 0.0405 \;,
\quad 0.021 < m_u/m_s < 0.028 \;,
\end{equation}
giving $3.08 < m_u < 4.15$~MeV.
These light quark masses and their ratios are consistent with
those obtained in Refs.~\cite{mass,ratio},
but do not agree as well with some other recent studies~\cite{donoghueud},
in which $m_u/m_d \alt  0.3$ was obtained.

The heavy quark masses are now constrained to the narrow ranges
\begin{equation}
1.19 < m_c < 1.23 \;, \qquad 4.09 < m_b < 4.20\;.
\end{equation}
Another interesting result is restrictive ranges
for the constants $B_K$ and $f_B$
\begin{equation}
0.33 < B_K < 0.43 \;, \qquad 0.14 < \sqrt{B_B\,} f_B < 0.17~{\rm GeV} \;,
\end{equation}
on which theoretical uncertainties have been problematic~\cite{geng}.

We conclude with some brief remarks.
{}From Eq.~(\ref{s-d}), $m_t$ is inversely proportional to
$\eta_b\eta_c\eta^{1/2}$, and the theoretical uncertainty in this quantity
could somewhat enlarge or close the window in $m_t$ (and correspondingly the
window in $|V_{cb}|$).  The DHR analysis assumes dominance of the top quark
Yukawa couplings in the RGE evolution. Since the output $\tan\beta$ may be
large, the effects of fully including $\lambda _b$ and $\lambda _{\tau}$ in the
evolution may not be negligible; this  question deserves further study. Also
two-loop renormalization group equations between $M_Z$ and $M_{\rm GUT}$ should
eventually be incorporated. We have studied the charged Higgs boson effects on
$\epsilon_K^{}$ and $r_d$. With $M_{H^{\pm}}$  degenerate with $m_t$, as
assumed in the model, we found no significant changes in our results. This is
due to the fact that $H^\pm$ effects are smaller at large $\tan\beta$ for the
$K$ and $B$ systems. In summary, the DHR ansatz for fermion mass matrices is
consistent with all current experimental constraints at 90\%~C.L.. It leads to
almost unique values for $m_t$ and quark mixings which make it an interesting
target for future experiments.

\acknowledgements
We thank M.~Barnett and K.~Hikasa for advance information from the Particle
Data Group, G.~Anderson and L.~Hall for communications, and A.~Manohar
for comments on quark mass ratios.
This work was  supported in part by the U.S.~Department of Energy under
contract
No.~DE-AC02-76ER00881 and in part by the University of Wisconsin
Research Committee with funds granted by the Wisconsin Alumni Research
Foundation. T.~Han was supported by an SSC Fellowship from the Texas
National Research Laboratory Commission under Award No. FCFY9116.

\figure{\label{scatter}
Scatter plot of $\sin\beta$ versus $m_t$ from our Monte Carlo analysis
of the DHR model, imposing the constraints of input masses and present values
of CKM matrix elements.}

\end{document}